# On Deuteron Spin Dependent Structure Function $g_2(x)$ in Relativistic Model of Deuteron. *

N.L.Ter-Isaakyan

*Yerevan Physics Institute, Republic of Armenia*
(Alikhanian Brothers St. 2, Yerevan 375036, Armenia)
e-mail: norair@vx1.yerphi.am

## Abstract

The spin dependent deuteron structure functions are investigated in the relativistic model of deuteron. It is shown that the deuteron structure function $g_2(X)$ can be considered properly only in the infinite momentum frame where the virtual photon has pure transverse components ($q_0 = -q_3 \to 0$ at $P \to \infty$), whereas in the conventional Breit frame the impulse approximation for $g_2(X)$ may be violated due to $q\bar{q}$ pair creation by the virtual photon. It is shown that due to relativistic effects the deuteron structure function $g_2(X)$ has a nontrivial contribution determined by first derivative of the nucleon structure function $g_1(x)$. Such contribution is small but it increases when $X \to 1$. It may achieve about 10 % at $X \sim 0.75$ and could be essentially larger for the transverse structure function $g_\perp(X)$.

## 1 Introduction

The investigation of the spin-dependent nucleon structure functions (SF) has attracted much attention recently. Experiments are prepared at SLAC(E142/143) [1], CERN(SMS) [2] and DESY(HERMES) [3]. The neutron SF can be extracted from the light nuclei ($^2$D or $^3$He) data, and nuclear effects should be taken into account to obtain accurate results. The procedure involves convolution of the nucleon infinite momentum frame (IMF) (or light cone) distributions in nucleus with parton distributions in the nucleon [4, 5]. On the other hand, the structure function $g_2(x)$ has no simple interpretation in probabilistic language [7, 8, 9] and it is natural to expect that the procedure of extracting $g_2^n(x)$ from nuclear data should be more complicated.

*Work supported in part by International Science Foundation, Grant ♯ RYE OOO and by the Project INTAS 93-283.

The nuclear effects for spin dependent structure functions were investigated in a number of papers (see, e.g. [5, 6, 12, 13] and references therein) in various model approaches and it was shown that relativistic effects could violate the impulse approximation. We investigate this question in the framework of Frankfurt-Strikman relativistic deuteron model [4]. We consider a deuterium target, but, qualitatively, our results are valid for any few-nucleon nuclei.

We show that even in the IMF the $q\bar{q}$-pairs creation or annihilation by electromagnetic field could violate impulse approximation for $g_2^D(x)$. Following the arguments of our previous paper [9], where in the special IMF a proper interpretation of $g_2(x)$ in terms of quark-parton wave functions has been proposed, we find that in such system the deuteron structure function $g_2^D(x)$ can be expressed through nucleon structure functions by some generalized convolution equation. We find that $g_2^D(x)$ is determined not only by $g_2^N(x)$ but also by derivative of $xg_1^N(x)$. The second term arises due to the nucleon internal motion and vanishes for noninteracting nucleons. The numerical estimates show that corresponding contribution is small but increases strongly at $X \to 1$ and may achieve about 10% at $X \sim 0.75$. For the transverse structure function $g_\perp^D(x)$ the role of such term strongly depends on the twist-3 contribution and could be considerably larger.

## 2 Kinematics. Impulse Approximation

The electron deep-inelastic scattering on spin-one target can be described by the hadronic tensor:

$$\frac{M}{4\pi}W_{\mu\nu}(\lambda',\lambda;P,q) = \sum_X (2\pi)^4 \langle P,\lambda'|J_\mu|X\rangle\langle X|J_\nu|P,\lambda\rangle\delta(P+q-P_X), \qquad (1)$$

where $P$ and $q$ are deuteron and virtual photon 4-momenta, $\lambda$ and $\lambda'$ are components of initial and final deuteron spins along $z$-axis, $J_\mu$ is the electromagnetic current. For spin-one target the electromagnetic tensor depends on eight invariant structure functions [10]. We are interisted only in $g_1(x)$ and $g_2(x)$, which are determined through antisymmetric part of hadronic tensor in the same form as for the spin 1/2 target:

$$\frac{M}{4\pi}W_{\mu\nu}^a(\lambda',\lambda;P,q) = iM\epsilon_{\mu\nu\lambda\sigma}q^\lambda\{M^2 s^\sigma G_1 + (P\cdot q s^\sigma - s\cdot q P^\sigma)G_2\}, \qquad (2)$$

where the 4-vector $s^\mu$ characterizes the target polarization state and is expressed in the following form

$$s^\mu = -\frac{iP_\nu}{M}\epsilon^{\nu\mu\alpha\beta}\varphi_{*\alpha}^{\lambda'}(P)\varphi_\beta^\lambda(P) \qquad (3)$$

through initial and final deuteron polarization vectors $\varphi_{*\alpha}^{\lambda'}(P)$ and $\varphi_\beta^\lambda(P)$

$$P^\mu\varphi_\mu(P) = 0, \quad \varphi_\mu^*(P)\varphi^\mu(P) = -1. \qquad (4)$$

In terms of old fashioned perturbation theory the impulse approximation for the hadronic tensor (1) corresponds to the diagram of fig.1a, which is supposed to dominate in IMF, the diagrams with $q\bar{q}$-pairs creation or annihilation by electromagnetic current (fig.1b) being suppressed at $P \to \infty$. But this argument is not valid for $g_2(x)$. Let us demonstrate this statement in the conventional frame, where photon has pure z-component [11]:

$$P_\mu = (E, P, 0, 0), \quad q_\mu = (0, -2Px, 0, 0). \tag{5}$$

In this system spin dependent SF can be expressed through antisymmetric parts of $W^a_{ij}$ and $W^a_{i0}$ ($i = 1, 2$) as follows:

$$\frac{1}{2\pi} W^a_{ij} = 2i\epsilon_{ij} g_1(x) \frac{s_0}{2P}$$

$$\frac{1}{2\pi} W^a_{i0} = i\epsilon_{ij} s_j (g_1(x) + g_2(x)) \frac{2M}{P} \tag{6}$$

where the functions $g_1 = M^2 \nu G_1$ and $g_2 = M\nu^2 G_2$ scale in the Bjorken limit, $M$ is the deuteron mass, $\nu = P \cdot q/M$, $x = Q^2/2P \cdot q$; we also use more conventional variable $X = 2x$ related to one nucleon in the deuteron.

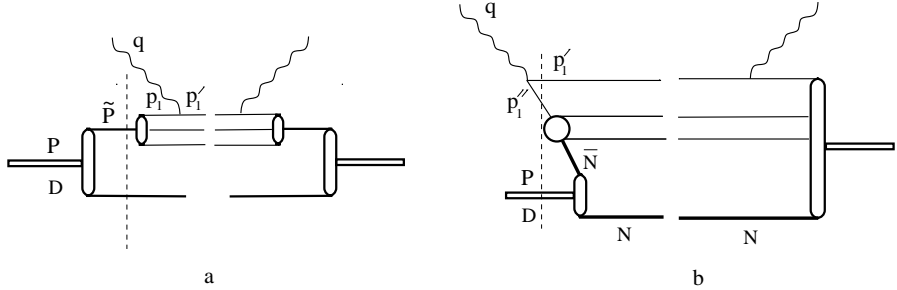

**fig. 1**: Deep inelastic scattering on deuteron: a) diagram corresponding to impulse approximation, b) diagram which could violate impulse approximation.

For the quark momenta (which are defined on diagrams) we introduce standard parameterizations:

$$\vec{\tilde{P}} = \eta \vec{P} + \vec{k}_\perp \quad \vec{k}_\perp \vec{P} = 0$$

$$\vec{p}_1 = x_1 \vec{P} + \vec{p}_{1\perp}, \quad \vec{p}'_1 = -x_1 \vec{P} + \vec{p}_{1\perp}, \quad \vec{p}''_1 = -x_1 \vec{P} - \vec{p}_{1\perp},$$

$$\vec{p}_{1\perp} \vec{P} = 0. \tag{7}$$

The vertices of photon interactions with quarks (on fig.1a) and with $q\bar{q}$-pairs (on fig.1b) at $P \to \infty$ behave as follows ($i,j=1,2$):

$$\bar{u}(p'_1)\gamma_i u(p_1) = 2Px_1 \sigma_i \sigma_3, \quad \bar{u}(p'_1)\gamma_0 u(p_1) = 2(m + i\epsilon_{ik}\sigma_i p_{1k}),$$

$$\bar{u}(p'_1)\gamma_i v(p''_1) = 2(m\sigma_i + i\epsilon_{ij} p_{1j})\sigma_2, \quad \bar{u}(p'_1)\gamma_0 v(p''_1) = 2Px_1 \sigma_3 \sigma_2. \tag{8}$$

The amplitude of antinucleon interaction with deuteron on fig.1b is $\sim P$. Hence, the large energy denominators corresponding to dashed lines on diagrams of fig.1b may be compensated for $W_{i0}$ (but not for $W_{ij}$). These diagrams may contribute to $(g_1(x)+g_2(x))$ and, therefore, may violate the impulse approximation.

Let us consider now IMF where photon have pure transverse component at $P \to \infty$:

$$P_\mu = (E, P, 0, 0), \quad q_\mu = (\frac{q_\perp^2}{4Px}, \frac{-q_\perp^2}{4Px}, \vec{q}_\perp). \tag{9}$$

In this frame the structure functions $g_1(x)$ and $g_2(x)$ are expressed through antisymmetric components of the hadronic tensor $W_{i0}^a$ ($i = 1, 2$) as follows:

$$\frac{q_\perp^2}{4\pi Px} W_{i0}^a(\lambda', \lambda) = 2i\epsilon_{ij}q_j \{g_1(x)\lambda\delta_{\lambda'\lambda} + \frac{2Mx}{q_\perp^2} g_2(x)(\vec{s}_\perp\vec{q}_\perp)_{\lambda'\lambda}\} \tag{10}$$

The structure function $g_2(x)$ can be extracted from the spin-flip amplitude, ($\lambda' = \pm 1$, $\lambda = 0$), $s_0 = 0$, $\vec{s}_\perp\vec{q}_\perp = 1/\sqrt{2}(q_x - i\lambda q_y)$. In the coordinate system (9) the quark-photon vertices behave as follows:

$$\bar{u}(p_1')\gamma_0 u(p_1) = 2Px_1, \quad \bar{u}(p_1')\gamma_i u(p_1) = 2(p_{1i} + q_i + i\epsilon_{ik}q_k\sigma_3),$$
$$\bar{u}(p_1')\gamma_0 v(p_1'') = \vec{\sigma}\vec{q}\sigma_3\sigma_2, \quad \bar{u}(p_1')\gamma_i v(p_1'') = -2Px_1\sigma_i\sigma_2. \tag{11}$$

The contribution of fig.1b diagrams does not vanish at $P \to \infty$. But the corresponding values of spin-flip amplitude $W_{0i}^a$ for such diagrams are independent on $\vec{q}$ direction and therefore do not contribute to $g_2(x)$. It is easy to see if one compare terms proportional to $q_x q_y$ at both sides of (10). For the spin-nonflip part of the hadronic tensor, which determines $g_1(x)$ these diagrams also do not contribute in Bjorken limit.

Having been convinced that the diagrams with $\bar{q}q$ pair creation do not contribute to $g_1^D(x)$ and $g_2^D(x)$, we do not need to invoke their parton representation and derive the impulse approximation immediately for the structure functions. According to fig.1a diagram, we can derive in IMF (9) the deuteron hadronic tensor through nucleon one:

$$W_{\mu,\nu}^D(P, \lambda', \lambda) = \sum_{N, s_1, s_1', s_2} \int \frac{d\Gamma}{\eta} \Psi_{s_1, s_2}^{*\lambda'} \Psi_{s_1', s_2}^{\lambda} W_{\mu,\nu}^N(\tilde{P}, s_1, s_1'), \tag{12}$$

where the sum is running over $N = p, n$ and $d\Gamma$ is two particle phase space:

$$d\Gamma = \frac{d^2 k_\perp d\eta}{(2\pi)^3 2\eta(1 - \eta)}. \tag{13}$$

Note, that we neglect the admixture of pions, gluons, or six quark states in the deuteron IMF wave function.

The energy denominator, corresponding to the dashed line on the fig.1a diagram is included into the definition of deuteron WF:

$$\psi_{s_1, s_2}^\lambda(\eta, \vec{k}_\perp) = \frac{\Gamma_{s_1, s_2}^\lambda(\eta, \vec{k}_\perp)}{2P(E - E_1 - E_2)} \tag{14}$$

where $\Gamma_{s_1, s_2}^\lambda$ is the deuteron-nucleon vertex function.

The deuteron and nucleon Bjorken variables $x = q_\perp^2/2P \cdot q$, and $\tilde{x} = q_\perp^2/2\tilde{P}q$ in IMF (9) are connected as follows:

$$x = \frac{\tilde{x}\eta}{1 + \frac{2\vec{k}_\perp \vec{q}_\perp}{q_\perp^2}\tilde{x}}. \tag{15}$$

We emphasize the term $2\vec{k}_\perp \vec{q}_\perp / q_\perp^2$ in (15) is kept in the scaling limit. Later we shall see that such term defines nontrivial contribution to $g_2^D(x)$.

Equation (13) determines the impulse approximation. Similar equation has been actually obtained in [5] but in terms of probabilities and therefore cannot be immediately used for extraction of $g_2^D(x)$.

The following circumstance should be noted. The equation (13) is not an exact equation even in the special IMF (9). It could be violated for spin-flip amplitudes, and should be understood only under definite prescriptions. For the antisymmetric part of $W_{0i}$ it is valid, for instance, only in terms proportional to $q_x q_y$.

The nucleon in our approach is off energy shell and its structure functions may depend on its off-shellness, i.e. on $k_\perp^2/m_\perp^2$ and on $1 - 2\eta$. The off-shell effects were investigated in [12, 13]. We ignore such dependence and take into account only relativistic corrections which arise due to proper definition of deuteron IMF WF.

## 3  The deuteron IMF wave function

For numerical calculations the deuteron IMF wave function which is defined according to Eq.(14) should be connected with phenomenological rest frame WF. We do not discuss different approaches (see [13, 14] and references therein) and follow Frankfurt-Strikman prescription [4].

It is convinient to write the spin-orbital part of deuteron WF via Melosh matrices, which determine their transformations from rest frame into IMF. [15, 16].

$$\Psi_{s_1,s_2}^\lambda(\eta, \vec{k}_\perp) = \sum_{s_1',s_2'} \Phi(M_0^2) \, U_{s_1,s_1'}(\eta, \vec{k}_\perp) U_{s_2,s_2'}(1-\eta, -\vec{k}_\perp) \, \psi_{s_1',s_2'}^\lambda(\eta, \vec{k}_\perp) \tag{16}$$

where $\Phi(M_0^2)$ is radial part of deuteron WF which is supposed to depend only on one argument- invariant mass of two-nucleon system [4]

$$M_0^2 = \frac{k_\perp^2 + m^2}{\eta(1-\eta)}. \tag{17}$$

where $m$ stands for nucleon mass. The Melosh matrix has a following form:

$$U(\eta, \vec{k}_\perp) = \frac{m + M_0 \eta + i\epsilon_{ik}\sigma_i k_\perp^k}{\sqrt{(m + M_0 \eta)^2 + k_\perp^2}} \tag{18}$$

The spin-orbital parts of deuteron rest frame WF for S- and D-waves we write via Pauli spinors $w_\alpha^s$:

$$\psi_{s_1 s_2}^{u\lambda} = \frac{e_i^\lambda w^{*s_1} \sigma_i \sigma_2 w_2^s}{\sqrt{2}},$$

$$\psi_{s_1 s_2}^{w\lambda} = \frac{e_i^\lambda w^{*s_1} \sigma_j \sigma_2 w_2^s}{2} \left( \frac{3k_i k_j}{\vec{k}^2} - \delta_{ij} \right), \tag{19}$$

where $e_i^\lambda$ are deuteron rest frame polarization vectors, $\vec{k} = (k_3, \vec{k}_\perp)$, $k_3 = (\eta - \frac{1}{2}) M_0$ is the nucleon 3-momenta in the two-nucleon rest frame.

The normalization of deuteron WF (14) can be fixed from electric or baryon charge conservation:

$$\int \Psi^{*\lambda} \Psi^\lambda \, d\Gamma = 1. \qquad (20)$$

The deuteron IMF wave functions defined according to (14), (16) and (19) coincide with Frankfurt-Strikman WF ([4, 5]), but is written in a more compact form. To connect our WF (14) with existing parametrizations, we introduce conventionally normalized wave functions:

$$U(k) = \frac{1}{\pi} \sqrt{\frac{\vec{k}^2}{2\omega}} \, \Phi_u(M_0^2), \quad W(k) = \frac{1}{\pi} \sqrt{\frac{\vec{k}^2}{2\omega}} \, \Phi_w(M_0^2)$$

$$\int (U^2 + W^2) dk = 1. \qquad (21)$$

# 4 Generalized convolution equation for $g_1(x)$ and $g_2(x)$

To derive the deuteron structure function consider Eq.12 for antisymmetric part of $W_{i0}^a$ and compare terms proportional to $q_x q_y$. Making use of Eq.(10) and analogous equation for the nucleon hadronic tensor we find in terms of deuteron IMF wave functions (16):

$$\lambda g_1^D(x) = \int \frac{d\Gamma}{\eta} \Psi^{*\lambda} \sigma_3 \Psi^\lambda \, g_1^N(\tilde{x}), \qquad (22)$$

$$\frac{q_x - i\lambda q_y}{\sqrt{2}} \, 2M g_2^D(x) = \int \frac{d\Gamma}{\eta} \{\Psi^{*\lambda}(\vec{\sigma}\vec{q})\Psi^0 \frac{2m}{\eta} \, g_2^N(\tilde{x}) + \Psi^{*\lambda} \sigma_3 \Psi^0 \, \frac{\eta q_\perp^2}{x^2} \, \tilde{x} g_1^N(\tilde{x})\}, \qquad (23)$$

where $g_i^N(X) = g_i^p(X) + g_i^N(X)$. The second term in (23) at first sight seems do not contribute to $g_2^D(x)$, because the value of $\Psi^{*\lambda} \sigma_3 \Psi^0$ is proportional to linear power of the nucleon transverse momenta $\sim (k_x - i\lambda k_y)$ and vanishes after taking integration on $\vec{k}_\perp$. But if we take the exact equation (15) connecting deuteron and nucleon Bjorken variables and expand $\tilde{x} g_1^N(\tilde{x})$ at $q_\perp^2 \to \infty$ keeping terms $\sim \vec{k}_\perp \vec{q}_\perp / q_\perp^2$, we find nonvanishing contribution of this term determined by first derivatives of $\tilde{x} g_1^N(\tilde{x})$.

Note, that relativistic corrections are expected to be small $(k_\perp^2/m^2) \sim 1/200$) and we neglect terms proportional to $k_\perp^2/m^2$ as compared to unity but keep them where they are enhanced by a factor $1/(1-X)$ at $X \to 1$. We also keep terms proportional to $k_3/m$. The neglected terms are essential in the range kinematically forbidden for free nucleon ($X > 1$). If necessary, they can be easily derived from Eqs.(22),(23). We present the simplified equations:

$$g_1^D(x) = \int_x^1 \frac{d\Gamma}{\eta} \, (\Phi_u - \frac{\Phi_w}{\sqrt{2}}) \, [(\Phi_u + \frac{\Phi_w}{\sqrt{2}}) - \sqrt{2}\Phi_w \frac{k_3}{m}] \, g_1^N(\tilde{x}) \qquad (24)$$

$$g_2^D(x) = \int_x^1 \frac{d\Gamma}{\eta} \frac{m}{M\eta} \, (\Phi_u - \frac{\Phi_w}{\sqrt{2}}) \, \{[(\Phi_u + \frac{\Phi_w}{\sqrt{2}}) - \frac{\Phi_w}{\sqrt{2}} \frac{k_3}{m}] \, g_2^N(\tilde{x}) +$$

$$[-\frac{k_3}{\sqrt{2}m}\Phi_w + \frac{k_\perp^2}{2m^2}\Phi_u] \frac{d}{d\tilde{x}} [\tilde{x}g_1^N(\tilde{x})]\} \quad (25)$$

where $\Phi_u(M_0^2)$ and $\Phi_w(M_0^2)$ are $S$- and $D$- wave radial wave functions. Note, that $g_1^N$ contribution to $g_2^D(x)$ arises due to $D$-wave part of deuteron wave function (the first term in the square brackets in (25)) and in $S$-wave part due to Melosh transformation (second terms). Second term in square brackets in (24) is also originated from Melosh matrices. Eq.24 coincides with results of Ref.([5]).

It is easy to see that for $g_2^D(x)$ Eq.(25) satisfies the Burkhard-Cottingham sum rule [17]

$$\int_0^1 g_2(x)\ dx = 0. \quad (26)$$

Eqs.(24) and (25) can be considered as a generalization of naive convolution equation. Due to relativistic effects the deuteron spin dependent SF cannot be presented in pure probabilistic form in terms of deuteron rest frame WF. Together with $g_1^N(x)$ contribution to $g_2^D(x)$, due to Melosh transformation a nondiagonal terms involving the interference of $S$- and $D$-wave part of the deuteron WF also contribute to deuteron SF. Note also an extra factor of $1/\eta$ in (25).

It is well known that $g_2(x)$ vanishes for noninteracting constituents. In that sense Eq.(25) has a transparent interpretation. The first term corresponds to quark gluon interactions forming a nucleon and vanishes for noninteracting partons, second term corresponds to nucleon interaction forming a deuteron and vanishes for noninteracting nucleons.

Our results, in general, resemble results of Refs.[12, 13], where the binding effects were investigated in a covariant framework. But it is difficult to establish a detailed connection between our aproach and that of [12, 13], because underlying model assumption are different.

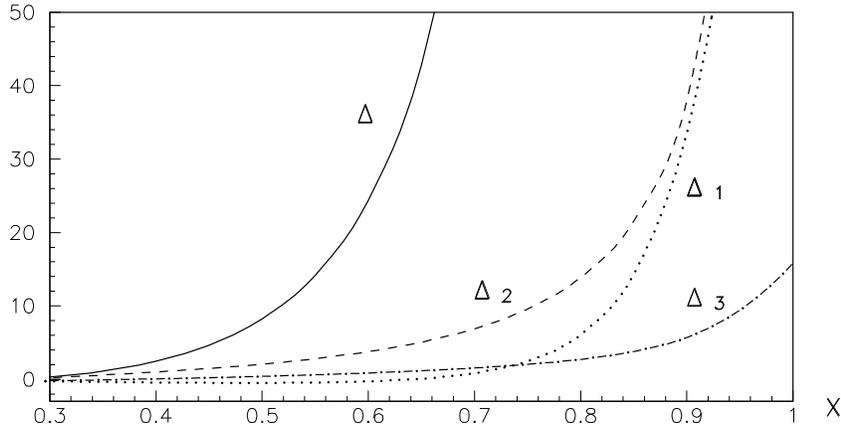

**fig.** 2: Relativistic corrections to $g_2^D(X)$, $\Delta_1(X)$, $\Delta_2(X)$, $\Delta_3(X)$ and to $g_\perp^D(X)$, $\Delta(X)$ (in %).

In Ref.[6] the nuclear effects for $g_2^D(x)$ were investigated in the framework of operator product expansion on light cone. The authors have considered a model of nucleon binding by a pion field. It is shown that impulse approximation for $g_2^D(x)$ is violated by nondiagonal contributions of states with different number of pions in the initial and final deuteron wave function. Keeping in mind that the space-time picture of bound state depends on the coordinate system, a following connection of our result with result of Ref.[6] can be established. In the IMF (9) a part of NN$\pi$ interaction which brought to nondiagonal contributions in [6] are hidden, in our approach, in the deuteron wave function and brought to second term in (25). The remaining part of interaction should be taken into account explicitly in terms of pion admixture in deuteron wave function. It seems that we take into account the essential part of NN interaction which can be approximately treated as potential, and Egs.(24) and (25) give good generalization of convolution equations.

We present numerical estimates only for illustration. Relativistic effects turn out to be small but increases considerably at $X \to 1$. According to model calculations [18] the twist-3 contribution to $g_2^N(x)$ at $x > 0.5$ is, possibly, small. In numerical estimates we take for simplicity $g_1^N(x) \sim (1-x)^3$ and define $g_2^N(x)$ through Wandzura-Wilchek relation [19]. We perform calculations with deuteron WF of the Paris NN potential [20]. The relative contributions of different terms in Eqs.(24) and (25) are presented on fig.2; $\Delta_1(X)$ is smearing correction to $g_1^D$ (arises in first term of (24)), $\Delta_2(X)$ is relative contribution of $g_1^N(\tilde{x})$ to $g_2^D(X)$, $\Delta_3(X)$ is relative contribution of the second term in (24), which arise due to Melosh transformation. At $X \sim 1$ this quantities behave as follows:

$$\Delta_1(X) \sim \Delta_2(X) \sim \frac{3k_\perp^2}{2m^2} \frac{X^2}{(1-X)^2}, \qquad \Delta_3(X) \sim \frac{3k_\perp^2}{4m^2} \frac{X}{1-X}. \qquad (27)$$

Thus at $X \sim 1$ the $g_1^N(\tilde{x})$ contribution to $g_2^D(X)$ turns out of the same order than typical smearing correction and at lower values of $X$ even exceeds it. It becomes noticeable at $X \sim 0.5$ ($\sim 2\%$) and achieves 10% at $x \sim 0.75$.

The structure function $g_2(x)$ can be determined by measuring transverse asymmetry, which defines transverse structure function $g_\perp(x) = g_1(x) + g_2(x)$, and the role of this correction is more important for $g_\perp(x)$. On fig.2 we also present the relative contribution of second term in (25) to $g_\perp^D(X)$, $\Delta(X)$, arising from $(\tilde{x}g_1^N(\tilde{x}))'$. If twist-3 contribution is really small, $g_\perp^N(\tilde{x})$, will be small as compared with $g_1^N(\tilde{x})$ and $\Delta(X)$ could considerably exceed $\Delta_2(X)$. In our example we neglect the twist-3 part of $g_2^N(\tilde{x})$ and $\Delta(X)$ turned out too large. Nevertheless it is clear that this correction could be very important in extracting $g_2^n(x)$ from deuteron data.

## 5  Acknowledgements

A part of this work was performed at DESY. I would like to thank DESY Hermes group for kind hospitality. I thank A.Airapetian and V.Garibian for help in computer calculations.